\documentclass[%
 aip,
 jmp,%
 amsmath,amssymb,
 reprint,%
]{revtex4-1}

\usepackage{graphicx}
\usepackage{dcolumn}
\usepackage{bm}

\begin{document}

\title{Coupling between quantum Hall state and electromechanics in suspended graphene resonator}

\author{Vibhor~Singh}
\email{vibhor@tifr.res.in}
\author{Bushra~Irfan}
\author{Ganesh~Subramanian}
\author{Hari~S.~Solanki}
\author{Shamashis~Sengupta}
\author{Sudipta~Dubey}
\author{Anil~Kumar}
\author{S.~Ramakrishnan}
\author{Mandar~M.~Deshmukh}
\email{deshmukh@tifr.res.in}
\affiliation{Department of Condensed Matter Physics and Materials
Science, Tata Institute of Fundamental Research, Homi Bhabha Road,
Mumbai, 400005 India }


\begin{abstract}
Using graphene resonator, we perform electromechanical measurements in quantum Hall regime to probe the coupling between a quantum Hall (QH) system and its mechanical motion. Mechanically perturbing the QH state through resonance modifies the DC resistance of the system and results in a Fano-lineshape due to electronic interference. Magnetization of the system modifies the resonator's equilibrium position and effective stiffness leading to changes in resonant frequency. Our experiments show that there is an intimate coupling between the quantum Hall state and mechanics - electron transport is affected by physical motion and in turn the magnetization modifies the electromechanical response.
\end{abstract}

\maketitle

Nanoelectromechanical systems (NEMS) \cite{ekinci_review,CraigheadReview} have emerged as an active field for studying mechanical oscillations of nanoscale resonators and thereby provide a good platform for sensing mass \cite{akshaynaik,bachtoldmasssensing,Mass1b}, charge \cite{roukesElectrometer}, magnetic flux and magnetic moments \cite{bolle_observation_1999,davis_nanomechanical_2011,harris2DEG,HarrisRing}. Actuation and detection of these systems can be done using electrical signals; this makes them suitable for probing the coupling of electrical and mechanical properties \cite{park_C60,ZantNanotube,BachtoldNanotube}. An intriguing property of graphene is the anomalous quantum Hall effect (QHE) \cite{firstgraphene1,firstgraphene2}. Graphene NEMS have also been realized \cite{BunchScience,Chen1,resonator1,McEuenCVDresonator,chenRF}, and this enables us to conduct experiments to answer the following two questions - how is the conductance modified due to mechanical vibrations, and how does the quantum Hall (QH) state affect the electromechanics of the resonator? We find that magnetization of graphene plays a crucial role in providing a coupling to the mechanical motion.

In quantum Hall regime, graphene's density of states (DOS) spectrum splits into unequally spaced landau levels (LL). When Fermi energy lies between the gap of two LLs, the longitudinal resistance ($R_{xx}$) vanishes and transverse resistance ($R_{xy}$) gets quantized to $\frac{h}{\nu e^2}$, where $h$ is Planck's constant, $e$ is electronic charge and $\nu$ is called the filling factor given by the integers $\pm2, \pm6, \pm10...$ \cite{firstgraphene1,firstgraphene2}. Since $R_{xx}$ goes to zero, the quantization of $R_{xy}$ is reflected even in the two probe resistance. The gaps in DOS result in the oscillations in chemical potential with the magnetic field ($B$) for a fixed number of total charge carriers. This leads to the well known de Haas-van Alphen oscillations in magnetization \cite{kittel}. The ability to control the charge carrier density in graphene provides an additional knob to tune the magnetization of graphene in the quantum Hall limit.

Motivated by these ideas, we have probed the electromechanics of graphene resonators in the quantum Hall regime in very clean samples at low temperatures. We measure the two probe resistance of these devices, while mechanically perturbing them for different values of $B$. The system shows changes in the resistance upon mechanical perturbation and we provide a quantitative analysis for the observed Fano-lineshape in the resistance across the resonance. Further, we have looked at the complementary aspect, by measuring the resonant frequency ($f_0$) and quality factor ($Q$) of these resonators with $B$ and find that magnetization of graphene leads to a coupling to the mechanical motion. The effect of graphene's magnetization ($M$) gets enhanced at lower temperatures and we observe sharper response in the mechanical motion across the broken symmetry state of $\nu=1$.

\begin{figure}
\begin{center}
\includegraphics[width=80mm]{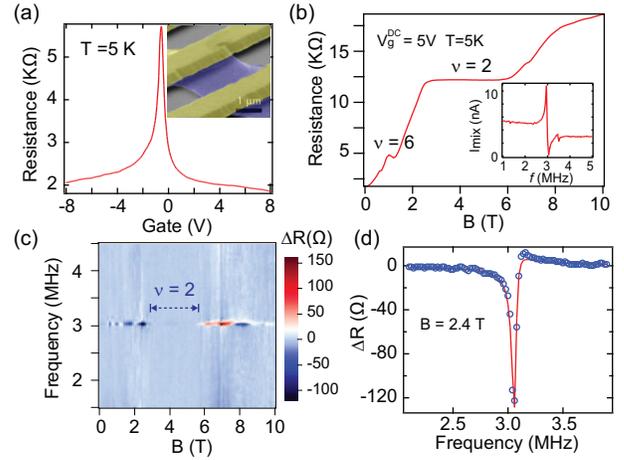}
\caption{\label{fig:rect} (Color online). (a) Two probe device resistance as a function of the gate voltage. 
Inset shows the scanning electron microscope image of a device with a scale bar of 1~$\mu$m. (b) Resistance as a function of magnetic field  at $V_g^{DC}$~=~5~V. Inset shows the measurement of the mixing current using heterodyne mixing technique to probe the resonant frequency of the resonator. (c) Colorscale plot of the change in resistance ($\Delta R$) with magnetic field and driving frequency. The resonant frequency of the mode is 3.05~MHz. (d) $\Delta R$ plotted with driving frequency along with the fitted curve using equation (1). The Fano-lineshape is clearly seen.}
\end{center}
\end{figure}

We fabricate the suspended graphene resonators using a well known process involving wet etching \cite{bolotinUltraclean,evaandrei-suspended}. The inset of Fig. \ref{fig:rect}(a) shows an image of one such device, where a monolayer graphene flake is suspended approximately 150~nm above the substrate. Fig. \ref{fig:rect}(a) shows the variation in resistance with gate voltage ($V_g^{DC}$) of a device at 5~K after current annealing \cite{bolotinUltraclean,evaandrei-suspended,bachtoldcleaning} which is possible due to the low contact resistance. The sharp Dirac peak at zero gate voltage and large mobility of charge carriers of $\approx$150,000 cm$^2$V$^{-1}$s$^{-1}$, are measures of the high quality of the sample; this is further confirmed in the magnetotransport measurements. Fig. \ref{fig:rect}(b) shows the two probe resistance measured as a function of the magnetic field. Different quantum Hall plateaus ($\nu=2, 6$), unique to the monolayer graphene, can be clearly seen.

Devices fabricated by this process show two kinds of resonant modes - resonant modes which arise from the suspended portion of gold electrodes that overlap with the graphene (gold modes) and modes arising from the mechanical motion of the graphene flake being clamped at two opposite edges (graphene modes) \cite{Chen1,resonator1}. At low temperatures for typical device dimensions, gold modes have resonant frequency of few MHz and do not show tunability with the gate voltage (data provided in supplemental material (SM) \cite{SM}). However graphene modes have larger resonant frequency ($\approx$100~MHz) and show tunability with gate voltage (see SM for tunability data). For electromechanical actuation and detection of the resonator, we have used the two-source heterodyne mixing \cite{SazanovaNature} and frequency modulation (FM) technique \cite{fmsmall}. In both the techniques, the detection of the mechanical motion relies on the finite transconductance $(\frac{dG}{dV_G^{DC}})$. Inset of Fig. \ref{fig:rect}(b) shows a measurement of a gold mode of the device using two source heterodyne mixing technique. The resonant frequency of this mode is very close to 3~MHz, which can be seen from the sharp change in the mixing current ($I_{mix}$).

In the first part of our experiment we probe the response of the quantum Hall state at resonant actuation. For this, we measure the DC resistance of the device while driving it through resonant frequency, by applying an RF signal at the gate electrode (see SM for details). The resonant frequency of the device can be measured using mixing technique prior to these measurements. Here, we focus on the response near the low frequency gold mode and discuss in detail the difference in response of the graphene mode later in the manuscript. We define change in resistance ($\Delta R$) by subtracting the device resistance measured away from the resonant frequency ($R_0(B)$) from the device resistance ($\Delta R(B)=R(B)-R_0(B)$). Fig. \ref{fig:rect}(c) shows $\Delta R$ as a function of driving frequency and magnetic field. Some features can be clearly observed : a) $\Delta R$ is zero in the plateau region, b) $\Delta R$ has different signs across the quantum Hall plateaus, and c) $\Delta R$ oscillates with $B$ for low values of magnetic field (see SM for detailed plot). To understand this, we model this geometry as a parallel plate capacitor, in which one plate (the substrate) remains fixed and the other plate (graphene flake) moves with large amplitude at the resonant frequency since it is tethered to the gold electrode. The voltage applied at the back gate influences the carrier density on the flake in two ways. Firstly, providing a driving force on the mechanical resonator leads to capacitance oscillations due to physical motion. Secondly, due to a direct capacitive coupling of RF voltage, the carrier density on the flake also oscillates. Within the adiabatic approximation, the total number of charge carriers ($N(t)$) on the flake at any time can be written as the sum of these two contributions. As the flake vibrates at the resonant frequency with large amplitude, $N(t)$ oscillates in time. Following this we can derive the expression for $\Delta R$ (see SM for detailed derivation) as:

\begin{align}\label{eqfano}
    \Delta R=\frac{d^2R}{{dV_g^{DC}}^2} \left( \frac{(\tilde{\omega}+q_x)^2+q_y^2}{\tilde{\omega}^2+1} -1 \right)(V_{ac}^2),
\end{align}

where, $q_x = -(\frac{1}{C_0}\frac{dC}{dz})(V_g^{DC}\frac{dF}{dV})(\frac{Q}{m \omega_0^2})$, $\tilde{\omega}=\frac{\omega-\omega_0}{\gamma/2}$ and
$q_y=1$.

Equation (\ref{eqfano}) agrees reasonably well with the observed variation of $\Delta R$ with magnetic field. Across the quantum Hall plateau in $R$, curvature $\frac{d^2R}{{dV_g^{DC}}^2}$ has opposite signs giving rise to negative and positive $\Delta R$ at the resonant frequency. Fig. \ref{fig:rect}(d) shows the line plot of $\Delta R$ from Fig. \ref{fig:rect}(c) at $B$~=~2.4~T and the fit using equation ($\ref{eqfano}$), which gives $q_x$~=~-4.18. This is close to our estimate of $q_x$~=~-4.80. During the quantum Hall plateau to plateau transition, we also see modification of Fano parameters ($q_{x,y}$) in equation (\ref{eqfano}) (see SM).
We would also like to contrast this with some high frequency graphene modes ($\approx$100~MHz) where we observe very small change in resistance with mechanical vibrations (see SM for detailed data). One reason for such a behavior might be the assumption that electron density in the flake follows the mechanical motion adiabatically at all frequencies; this may not be correct in magnetic field when charge density splits into compressible and incompressible regions. Such a time lag between drive and response would effectively reduce the value of $\Delta R$. The second possibility is that for these high frequency device modes, the spatial amplitude profile can be very complex \cite{bachtoldAFM,McEuenCVDresonator}. For the modes lofig:rectcalized at the edges, the ``effective density" which gets modulated by the mechanical motion can be small, and hence contributes little to $\Delta R$. Complex nature of these modes is also seen in the second aspect of our experiments that we discuss next.

\begin{figure}
\begin{center}
\includegraphics[width=60mm]{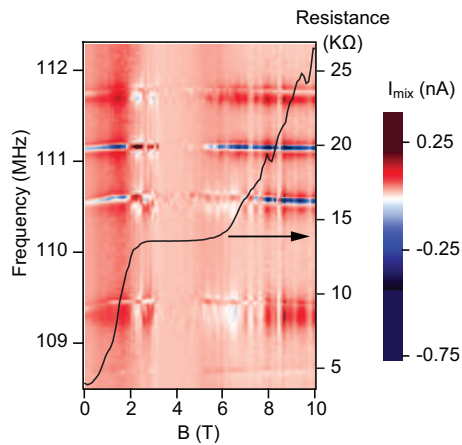}
\caption{\label{fig:multimode} (Color online). Measurement of mixing current using FM technique for multiple resonant modes with $B$ at $V_g^{DC}$~=~5~V and $T$~=~5~K. Closely spaced multiple modes can be clearly seen. The solid black line overlaid on the colorscale plot is resistance of the device at the same gate voltage and temperature.}
\end{center}
\end{figure}

The gold modes discussed above modulate a larger fraction of the total carrier density in the flake. However, low frequency and low quality factor for these modes reduce the sensitivity to the forces on the resonator. Therefore to look at the consequence of the quantum Hall state on the mechanics, namely the resonant frequency and quality factor, we turn our attention to the high frequency graphene modes with higher quality factor. At low temperatures, these devices show low tunability of resonant frequency with the gate voltage due to large built-in strain (see SM). Also, with a change in the gate voltage from zero, resonant frequency decreases \cite{Chen1,resonator1}. Such a change in resonant frequency at low temperature originates from the electrostatic softening of the spring constant \cite{resonator1}. Fig. \ref{fig:multimode} shows the measurement of $I_{mix}$ using FM technique for multiple modes as a function of magnetic field at $V_g^{DC}$~=~5~V. The presence of multiple modes in a very close range of frequency indicates that many of these are possibly edge modes arising due to the soft degrees of freedom along the length of the flake \cite{bachtoldAFM,McEuenCVDresonator} and their dispersion with $B$ is mode dependent. In Fig. \ref{fig:multimode}, the overlaid line plot shows resistance measured with increasing magnetic field at the same gate voltage. Clearly, we see larger $I_{mix}$ signal, when change in the resistance with magnetic field is large. However, in the plateau region, the resistance does not change with $V_g^{DC}$ ($\frac{dG}{dV_g^{DC}}\sim 0$), which makes the detection of resonant mode difficult in these regions. Surprisingly, for certain modes even in the plateau region, we can measure some detectable signal for mixing current (for example compare the modes at 110.5~MHz and 111.2~MHz) and it varies to a different extent across modes. These two observations suggest a mode dependent amplitude profile.

\begin{figure}
\begin{center}
\includegraphics[width=80mm]{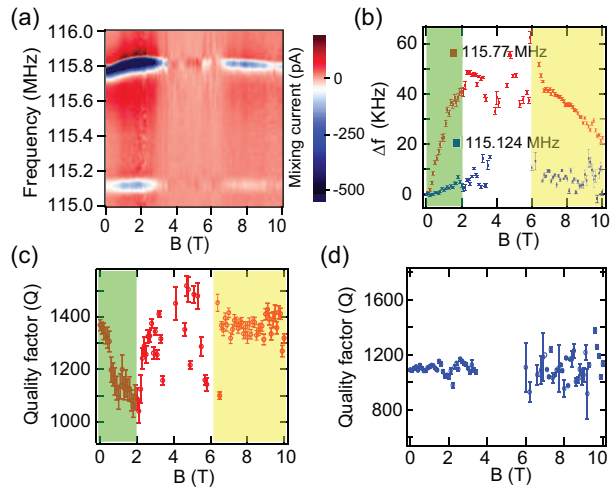}
\caption{\label{fig:figure5} (Color online). (a) Measurement of mixing current for two close-by mechanical modes with $B$ at 5~K and $V_g^{DC}$=~5~V. (b) Frequency shift ($\Delta f$) with $B$ by fitting the data in (a). Different behavior in frequency shift across the $\nu=2$ plateau is shown with shaded region. (c) and (d) show the variation of the quality factor with $B$ for the two modes shown in (a) at frequency 115.77~MHz and 115.124~MHz, respectively.}
\end{center}
\end{figure}

In Fig. \ref{fig:figure5}(a),  we have shown $I_{mix}$ for two modes, using FM technique, with magnetic field. It is evident from the colorscale plot that the two modes disperse differently with magnetic field. In Fig. \ref{fig:figure5}(b), the non-monotonic resonant frequency shift ($\Delta f$~=~$f_0(B)-f_0(0)$) with $B$ can be seen for the two modes. At low magnetic fields, for the upper mode (115.77~MHz) $\Delta f$ increases with $B$ accompanied by a reduction in $Q$ (green color shaded area in Fig. \ref{fig:figure5}(b) and (c)). At the $\nu=2$ plateau, $I_{mix}$ signal becomes very small ($\frac{dG}{dV_g^{DC}}\sim0$) and the estimation of $f_0$ and $Q$ becomes difficult. As $B$ is increased further, after the $\nu=2$ plateau, $\Delta f$ starts dropping slowly without much change in $Q$ (yellow color shaded area in Fig. \ref{fig:figure5}(b) and (c)). Similar behavior can be seen for the lower frequency mode (115.124~MHz), though the effect is less pronounced (Fig. \ref{fig:figure5}(d)).

\begin{figure}
\begin{center}
\includegraphics[width=80mm]{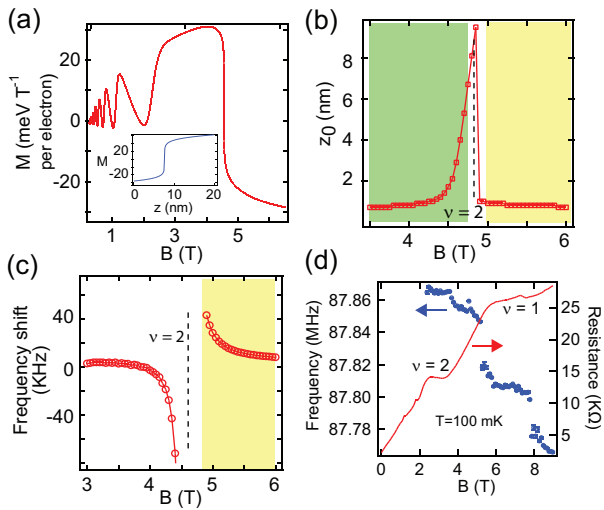}
\caption{\label{fig:calM4panel} (Color online). (a) Calculated magnetization ($M$) for model density of states with magnetic field ($B$). de Haas-van Alphen oscillations can be seen. Inset shows $M$ (with same units as shown in (a)) with the position of the flake ($z$) at $B$~=~4.8~T. (b) Equilibrium position of the flake ($z_0$) with $B$. Two different regimes across $\nu=2$ plateau are shaded with colors. (c) Change in resonant frequency with $B$ calculated from $k_{eff}$. (d) Measurement of resistance and resonant frequency at $T$~=~100~mK and $V_g^{DC}$~=~3~V.}
\end{center}
\end{figure}

To understand this, we examine various contributions to the total energy of the resonator ($E_{tot}$), which can be written as a sum of the mechanical energy, electrostatic energy and the contribution arising from the magnetization ($M$) of graphene ($E_{mag}$). The contribution of $E_{mag}$~($=-M.B$) to the total energy of the resonator comes from the magnetization's implicit dependence on $z$ through the total charge carriers ($N(z)$), where $z$ is the position of the flake from its equilibrium position and increases towards the gate. To understand the effect of magnetization of graphene on the total energy of the resonator, we calculate the magnetization of graphene by assuming a model DOS with LL width being the only free parameter. Result of such a calculation is plotted in Fig. \ref{fig:calM4panel}(a) (see SM for details about the calculations of $M$), where de Haas-van Alphen oscillations in $M$ can be clearly seen. Across the $\nu=2$ filling factor (close to $B$~=~4.2~T), $M$ shows a sharp jump. In the proximity of integer $\nu$, such a sharp change in $M$ makes it very sensitive to the position of the flake as shown in the inset of Fig. \ref{fig:calM4panel}(a). By minimizing the $E_{tot}$ with respect to $z$, the equilibrium position of the flake ($z_0$) with varying $B$ can be calculated and it is shown in Fig. \ref{fig:calM4panel}(b). It is clear that before the $\nu=2$ plateau as we increase the field, the flake moves closer to the substrate. Another important effect of $E_{mag}$ on the mechanics of the resonator is to modify the spring constant ($k_{eff}=\frac{d^2E_{tot}}{dz^2}|_{z=z_0}$) which leads to the frequency shift. In Fig. \ref{fig:calM4panel}(c), we show the change in resonant frequency calculated from $k_{eff}$ with varying $B$  (see SM for details about the calculations). From this model calculation, we can see that with an increase in $B$ on either side of the $\nu=2$ plateau, contribution of $E_{mag}$ leads to the softening of $k_{eff}$ and reduction in the resonant frequency. This is consistent with our experimental observation beyond $\nu=2$, where we observe a decrease in $\Delta f$ (yellow color shaded area in Fig. \ref{fig:figure5}(b) and Fig. \ref{fig:calM4panel}(c)).

However, before the $\nu=2$ plateau, we observe larger dissipation accompanying the increase in $\Delta f$ (green color shaded area in Fig. \ref{fig:figure5}(b) and (c)), which indicates that the contribution of $E_{mag}$ to $k_{eff}$ is not enough to fully understand such behavior and there are more subtle effects which arise when the flake moves closer to the substrate. Here, we note that enhancement of linear damping with magnetic field alone cannot explain the observed frequency shifts. The frequency shift due to linear damping ($\Delta f \approx - \frac{f}{8 Q^2}$) would be very small ($\approx$~-10~Hz, for the two modes), whereas we observe much larger positive $\Delta f$. It strongly suggests that there are other damping mechanisms present in the system. The basic characterization of the device in zero field shows that nonlinear damping mechanism \cite{bachtoldnonlinear} can become important as the flake moves closer to the substrate (see SM for detailed measurements). At lower values of magnetic field the frequency shift and decrease in the quality factor can be understood within this picture. With increasing $B$, as the flake moves closer to the gate it experiences enhanced nonlinear damping and a larger shift in frequency. From our numerical calculations for the coefficient of nonlinear damping, we get similar values as reported in ref \cite{bachtoldnonlinear} (details provided in SM).

For similar measurements on clean graphene devices at $T$~=~100~mK, the consequence of the chemical potential jumps in gapped regions of DOS gets enhanced. In Fig. \ref{fig:calM4panel}(d), we show the variation in resistance and the resonant frequency of a device with magnetic field. In resistance, the plateau at $\nu=2$  and $\nu=1$ can be clearly seen. The plateau at $\nu=1$ originates from the interaction induced broken symmetries of zeroth LL. Previous experiments seem to suggest that LL gap for $\nu=1$  is related to the broken valley degeneracy and not to the spin \cite{kim_high_field}. In the resonant frequency, we see jumps at certain fields over a monotonic decrease in resonant frequency. The jumps in resonant frequency can be thought of as arising from the magnetization changes as magnetic field sweeps the chemical potential across the gaps induced by the interactions. An overall decrease in resonant frequency and jump at $\nu=1$ plateau is beyond our model calculations and need further studies.

Our observation and quantitative analysis of resistance change giving a Fano-lineshape helps in better understanding the effect of electromechanical drive of graphene resonators. Also, the magnetization of graphene couples to the mechanical motion to modify the spring constant. The results can be further enhanced in devices with larger tunability of resonant frequency with gate voltage. We would like to emphasize that the change in the spring constant is a consequence of the coupling between magnetization and the carrier density set by the gate voltage. Therefore such results should be possible to see in other systems where the magnetization can be modified by both electrostatics and magnetic field.

The authors would like to thank J.~K.~Jain, Aashish~Clerk and Sunil~Patil for discussions and Padmalekha~K.~G. for help during the device fabrication. This work was supported by the Government of India.


\bibliographystyle{apsrev}


\end{document}